\documentclass[11pt,letter]{article}

\usepackage{amsmath}
\usepackage{amssymb}
\usepackage{amsthm}
\usepackage{fancyhdr}
\usepackage{makeidx}
\usepackage{url}
\usepackage{graphicx}
\usepackage{verbatim} 

\theoremstyle{definition}

\theoremstyle{definition}

\newcommand{\mycomment}[1]{}

\usepackage[left=1.25in,top=1.0in,right=1.25in,bottom=1.0in]{geometry}
\usepackage{amsfonts}
\usepackage{amssymb}
\usepackage{graphicx}
\usepackage{setspace}
\usepackage[round]{natbib}
\usepackage{amsmath}
\usepackage{amssymb}
\usepackage{amsbsy}
\usepackage{amsthm}
\usepackage{paralist}
\usepackage{bm}
\usepackage{booktabs}

\usepackage{mathtools}
\usepackage{cases}
\usepackage[boxed]{algorithm2e}
\usepackage{algpseudocode}

\newcommand{\N}{\mbox{N}}

\newcommand{\signif}{\phantom{}$^\star$}

\title{False discovery rate regression: an application to neural synchrony detection in primary visual cortex}
\author{James G.~Scott\footnote{University of Texas, Austin,
USA.  Correspondence to: james.scott@mccombs.utexas.edu}\\
Ryan C. Kelly\footnote{Google, New York, USA} \\
Matthew A. Smith\footnote{University of Pittsburgh, Pittsburgh, USA} \\
Pengcheng Zhou\footnote{Carnegie Mellon University, Pittsburgh, USA} \\
Robert E.~Kass\footnote{Carnegie Mellon University, Pittsburgh, USA}}

\date{First version: July 2013\\
This version: May 2014}

\begin{document}

\maketitle
\begin{abstract}
\noindent  Many approaches for multiple testing begin with the assumption that all tests in a given study should be combined into a global false-discovery-rate analysis.  But this may be inappropriate for many of today's large-scale screening problems, where auxiliary information about each test is often available, and where a combined analysis can lead to poorly calibrated error rates within different subsets of the experiment.  To address this issue, we introduce an approach called false-discovery-rate regression that directly uses this auxiliary information to inform the outcome of each test.   The method can be motivated by a two-groups model in which covariates are allowed to influence the local false discovery rate, or equivalently, the posterior probability that a given observation is a signal.   This poses many
subtle issues at the interface between inference and computation, and we investigate several variations of the overall approach.  Simulation evidence suggests that: (1) when covariate effects are present, FDR regression improves power for a fixed false-discovery rate; and (2) when covariate effects are absent, the method is robust, in the sense that it does not lead to inflated error rates.  We apply the method to neural recordings from primary visual cortex.  The goal is to detect pairs of neurons that exhibit fine-time-scale interactions, in the sense that they fire together more often than expected due to chance.  Our method detects roughly 50\% more synchronous pairs versus a standard FDR-controlling analysis.  The companion R package \verb|FDRreg| implements all methods described in the paper.

\end{abstract}

\newpage

\begin{spacing}{1.1}

\section{Introduction}

\subsection{Multiple testing in the presence of covariates}


The problem of multiple testing concerns a group of related null hypotheses $h_1, \ldots, h_n$ that are tested simultaneously.  In its simplest form, each test yields a summary statistic $z_i$, and the goal is to decide which of the $z_i$ are signals ($h_i = 1$) and which are null ($h_i = 0$).  Many solutions to this problem, such as Bonferroni correction, aim to control the family-wise error rate (FWER): the probability of incorrectly rejecting at least one null hypothesis, assuming that they are all true.  An alternative, which has become the dominant approach in many domains of application, is to control the false discovery rate (FDR): the proportion of false positives among those null hypotheses that are rejected \citep{benjamini1995}.  Regardless of which error rate they aim to control, however, most existing approaches obey a monotonicity property: if test statistic $z_i$ is declared significant, and $z_j$ is more extreme than $z_i$, then $z_j$ is also declared significant.  Yet in many cases, we have auxiliary covariate information about each test statistic, such as location in the brain or distance along a chromosome.  If significant test statistics tend to cluster in covariate space, then monotonicity becomes undesirable, and a procedure that takes account of the covariate should perform better.  In this paper, we introduce a method called \textit{false-discovery-rate regression} (FDRR) that incorporates covariates directly into the multiple-testing problem.

The method we describe here builds on the two-groups model \citep{efrontibshirani2001}, a popular framework for controlling the false-discovery rate.  In the two-groups model, some small fraction $c$ of the test statistics are assumed to come from an unknown signal population, and the remainder from a known null population.  Our proposal is to allow the mixing fraction $c$ to depend upon covariates, and to estimate the form of this dependence from the data.  Extensive simulation evidence shows that, by relaxing the monotonicity property in a data-dependent way, FDR regression can improve power while still controlling the global false-discovery rate.  The method is implemented in the publicly available R package \verb|FDRreg| \citep{scott:2014a}.

Our motivating application is the identification of interactions among many simultaneously recorded neurons, which has become a central issue in computational neuroscience.  Specifically, we use FDR regression to detect fine-time-scale neural interactions (``synchrony'') among 128 units (either single neurons or multi-unit groups) recorded simultaneously from the primary visual cortex (V1) of a rhesus macaque monkey \citep{kelly:etal:2010,kelly:kass:2012}.
The experiment from which the data are drawn produced thousands of pairs of neurons, each involving a single null hypothesis of no interaction.  In this case, combining all tests into a single FDR-controlling analysis would inappropriately ignore the known spatial and functional relationships among the neurons \citep[e.g.][]{smith:kohn:2008}.  Our approach for false-discovery rate regression avoids this problem: it detects roughly 50\% more significant neuron pairs compared with a standard analysis by exploiting the fact that spatially and functionally related neurons are more likely to exhibit synchronous firing.  

\subsection{The two-groups model}

In the two-groups model for multiple testing, one assumes that test statistics $z_1, \ldots, z_n$ arise from the mixture
\begin{equation}
\label{basicmixturemodel}
z \sim c \cdot f_1(z) + (1-c) \cdot f_0(z) \, ,
\end{equation}
where $c \in (0,1)$, and where $f_0$ and $f_1$ respectively describe the null ($h_i=0$) and alternative ($h_i=1$) distributions of the test statistics.  For each $z_i$, one then reports the quantity
\begin{equation}
\label{bayesoracle1}
 w_i = \mbox{P}(h_i = 1 \mid z_i) =  \frac{ c \cdot f_1(z_i)}{  c \cdot f_1(z_i) + (1- c) \cdot f_0(z_i) } \, .
\end{equation}
As \citet{efron:2008} observed, the information contained in $w_i$ provides a tidy methodological unification to the multiple-testing problem.  Bayesians may interpret $ w_i$ as the posterior probability that $z_i$ is a signal, while frequentists may interpret $1 - w_i$ as a local false-discovery rate.  The global false-discovery rate of some set $Z_1$ of putative signals can then estimated as
$$
\mbox{FDR}(Z_1) \approx \frac{1}{|Z_1|} \sum_{i: z_i \in Z_1} (1-w_i) \, .
$$
\citet{efrontibshirani2001} show that this Bayesian formulation of FDR is biased upward as an estimate of frequentist FDR, and therefore conservative.

An elegant property of the local FDR approach is that it is both frequentist and fully conditional: it yields valid error rates, yet also provides a measure of significance that depends on the precise value of $z_i$, and not merely its inclusion in a larger set \citep[c.f.][]{berger:2003}.  This can be achieved, moreover, at little computational cost.  To see how, observe that (\ref{bayesoracle1}) may be re-expressed in marginalized form as
\begin{equation}
1- {w}_i = \frac{(1-c) \cdot f_0(z_i)}{f(z_i)} \label{bayesoracle2} \, ,
\end{equation}
where $f(z) = c \cdot f_1(z_i) + (1-c) \cdot f_0(z_i)$ is the overall marginal density.  Importantly, $f(z)$ can be estimated from the empirical distribution of the test statistics; this is typically quite smooth, which makes estimating $f(z)$ notably easier than a generic density-estimation problem.  Therefore one may compute local FDR using cheap plug-in estimates $\hat{f}(z)$ and $\hat c$, and avoid the difficult deconvolution problem that would have to be solved in order to find $f_1(z)$ explicitly \citep[e.g.][]{efrontibshirani2001,newton:2002,martin:tokdar:2012}.

\subsection{FDR regression}

Implicit in the two-groups model is the assumption that all tests should be combined into a single analysis with a common mixing weight $c$ in (\ref{basicmixturemodel}).  Yet for some data sets, this may be highly dubious.  In our analysis of neural recordings, for example, a test statistic $z_i$ is a measure of pairwise synchrony in the firing rates of two neurons recorded from an array of electrodes, and these $z_i$'s exhibit spatial dependence across the array: two nearby neurons are more likely to fire synchronously than are two neurons at a great distance.  Similar considerations are likely to arise in many applications.

False-discovery-rate regression addresses this problem through a conceptually simple modification of (\ref{basicmixturemodel}), in which covariates $x_i$ may affect the prior probability that $z_i$ is a signal.  In its most general form, the model assumes that
\begin{eqnarray}
z_i &\sim& c(x_i) \cdot f_1(z_i) + \{1-c(x_i)\} \cdot f_0(z_i) \label{eqn:fdrr1} \\
c(x_i) &=& G\{s(x_i)\} \nonumber \,
\end{eqnarray}
for an unknown regression function $s(x)$ and known link function $G: \mathcal{R} \rightarrow (0,1)$.

This new model poses two main challenges versus the ordinary two-groups model (\ref{basicmixturemodel}).  First, we must estimate a regression model for an unobserved binary outcome: whether $z_i$ comes from $f_1$, and is therefore a signal.  Second, because each mixing weight in (\ref{eqn:fdrr1}) depends on $x_i$, there is no longer a common mixture distribution $f(z)$ for all the test statistics.  We therefore cannot express the Bayes probabilities in marginalized form (\ref{bayesoracle2}), and cannot avoid estimating $f_1(z)$ directly.

Our approach, described in detail in Section 2, is to represent $f_1(z)$ as a location mixture of the null density, here assumed to be a Gaussian distribution:
\begin{eqnarray*}
f_0(z) &=& \N(z \mid \mu, \sigma^2)  \\
f_1(z) &=& \int_{\mathcal{R}} \N(z \mid \mu + \theta, \sigma^2) \ \pi(\theta) \ d \theta \, . 
\end{eqnarray*}
Even in the absence of covariates, estimating the mixing density $\pi(\theta)$ is known to be a challenging problem, because Gaussian convolution heavily blurs out any peaks in the prior.  We consider two ways of proceeding.  The first is an empirical-Bayes method in which an initial plug-in estimate $\hat{\pi}(\theta)$ is fit via predictive recursion \citep{newton:2002}.  The regression function is then estimated by an expectation-maximization (EM) algorithm, treating $\hat \pi(\theta)$ as fixed.  The second is a fully Bayes method in which $\pi(\theta)$ and the regression function $s(x)$ are estimated jointly using Markov-chain Monte Carlo.  In simulation studies, both methods lead to better power and equally strong protection against false discoveries compared with traditional FDR-controlling approaches.

The rest of the paper proceeds as follows.  The remainder of Section 1 contains a brief review of the literature on multiple testing.  Section 2 describes both empirical-Bayes and fully Bayes methods for fitting the FDR regression model, and draws connections with existing approaches for controlling the false-discovery rate.  It also describes how existing methods for fitting an empirical null hypothesis may be combined with the new approach (Section \ref{sec:empirical_null}).  Section 3 shows the results of a simulation study that validates the frequentist performance of the method.  Section 4 provides background information on the neural sychrony-detection problem.   Section 5 shows the results of applying FDR regression to the sychrony-detection data set.  Section 6 contains discussion.

\subsection{Connection with existing work}

Our approach is based on the two-groups model, and therefore in the spirit of much previous work on Bayes and empirical-Bayes multiple testing, including \citet{efrontibshirani2001}, \citet{johnstone:silverman:2004}, \citet{scottberger06}, \citet{muller:etal:2006}, \citet{efron:2008,efron:2008b}, and \citet{bogdan:etal:2008}.  The final reference has a comprehensive bibliography.  We will make some of these connections more explicit when they arise in subsequent sections.

Other authors have considered the problem of multiple testing in the presence of correlation \cite[e.g.][]{clarke:hall:2009,fan:han:gu:2012}.  The focus there is on making the resulting conclusions robust to unknown correlation structure among the test statistics.  Because it explicitly uses covariates to inform the outcome of each test, FDR regression is different both in aim and execution from these approaches. 

On the computational side, we also draw upon a number of recent innovations.  Our empirical-Bayes approach uses predictive recursion, a fast and efficient method for estimating a mixing distribution \citep{newton:2002,tokdar:martin:ghosh:2009,martin:tokdar:2012}.  Our fully Bayes approach requires drawing posterior samples from a hierarchical logistic-regression model, for which we exploit the P\'olya-Gamma data-augmentation scheme introduced by \citet{polson:scott:windle:2012a}.

There is also a growing body of work on density regression, where an unknown probability distribution is allowed to change flexibly with covariates using nonparametric mixture models \citep[e.g.][]{dunson:pillai:park:2007}.  We do not attempt a comprehensive review of this literature, which has goals that are quite different from the present application.  For example, one of the key issues that arises in multiple testing is the need to limit the flexibility of the model so that the null and alternative hypotheses are identifiable.  Ensuring this property is not trivial; see \citet{martin:tokdar:2012}.  In density regression, on the other hand, only the overall density is of interest; the mixture components themselves are rarely identifiable.

Our application draws most directly on \citet{kelly:etal:2010} and \citet{kelly:kass:2012}.  We review other relevant neuroscience literature in Section 4.

\section{Fitting the FDR regression model} 

\subsection{An empirical-Bayes approach}

\label{sec:FDRregPR}

We use a version of the FDR regression model where
\begin{eqnarray}
z_i &\sim& c(x_i) \cdot f_1(z_i) +  \{1-c(x_i)\}  \cdot f_0(z_i) \nonumber \\
c(x_i) &=& \frac{1}{1+\exp\{-s(x_i)\}} \nonumber \\
f_0(z) &=& \N(z \mid \mu, \sigma^2) \nonumber \\
f_1(z) &=& \int_{\mathcal{R}} \N(z \mid \mu + \theta, \sigma^2) \ \pi(\theta) \ d \theta \, . \label{eqn:deconvolution1}
\end{eqnarray}
We have assumed a logistic link and a Gaussian error model, both of which could be modified to suit a different problem.  We also assume a linear regression where $s(x) = x^T \beta$, and therefore model non-linear functions by incorporating a flexible basis set into the covariate vector $x$.  Both $\mu$ and $\sigma^2$ are initially assumed to be known; in Section \ref{sec:empirical_null}, we describe how to weaken this assumption by estimating an empirical null, in the spirit of \citet{efron:2004}. 

The unknown parameters of the FDR regression model that must be estimated are the regression coefficients $\beta$  and the mixing distribution $\pi(\theta)$.  This section describes two methods---one empirical-Bayes, one fully Bayes---for doing so.  Both methods are implemented in the R package \verb|FDRreg|.

Our empirical-Bayes approach begins with a pre-computed plug-in estimate for $\pi(\theta)$ in (\ref{eqn:deconvolution1}), ignoring the covariates.  This is equivalent to assuming that $c(x_i) \equiv c$ for all $i$, albeit only for the purpose of estimating $\pi(\theta)$.  Many methods could be used for this purpose, including finite mixture models.  We recommend the predictive-recursion algorithm of \citet{newton:2002} for two reasons: speed, and the strong guarantees of accuracy proven by \citet{tokdar:martin:ghosh:2009}.  Predictive recursion generates a nonparametric estimate $\hat{\pi}(\theta)$, and therefore an estimate $\hat f_1(z)$ for the marginal density under the alternative, after a small number of passes (typically 5--10) through the data.  The algorithm itself is similar to stochastic gradient descent, and is reviewed in Appendix \ref{ap:PRdetails}.

\begin{algorithm}[t]
\KwData{Test statistics $z_1, \ldots, z_n$}
\KwIn{Densities $f_0(z)$, ${f}_1(z)$; initial guess $\beta^{(0)}$}
\KwOut{Estimated coefficients $\beta$ and posterior probabilities ${w}_i$}
\While{not converged}{
\begin{description}
\item[E step:] Update $Q(\beta) = E\{l(\beta) \mid \beta^{(t)} \}$ as
\begin{eqnarray*}
Q^{(t)}(\beta) &=& \sum_{i=1}^n \left\{ {w}_i^{(t)} x_i^T \beta - \log \left(1 + e^{x_i^T \beta} \right) \right\} \\
 w_i^{(t)} = E(h_i \mid \beta^{(t)}, z_i) &=&  \frac{c(x_i) \cdot  f_1(z_i)}{ c(x_i) \cdot  f_1(z_i) + \{1-c(x_i)\} \cdot f_0(z_i) } \\
c(x_i) &=&  \frac{1}{1+\exp\{ -x_i^T \beta^{(t)} \}} \, .
\end{eqnarray*}
\item[M step:] Update $\beta$ as
$$
\beta^{(t+1)} = \arg \max_{\beta \in \mathcal{R}^d} Q^{(t)}(\beta) \, 
$$
using the Newton--Raphson method.
\end{description}
}
\caption{\label{alg:emPR} EM for FDR regression using a plug-in estimate $\hat f_1(z)$.  To estimate $\hat{f}_1$, we use predictive recursion (Algorithm \ref{alg:PR}).}
\end{algorithm}

Upon fixing this empirical-Bayes estimate $\hat{f}_1(z)$, and assuming that $f_0(z)$ is known, we can fit the FDR regression model by expectation-maximization \citep{dempster:laird:rubin:1977}.  To carry this out, we introduce binary latent variables $h_i$ such that
\begin{eqnarray*}
z_i &\sim& 
\left\{ \begin{array}{l l}
f_1(z_i) & \mbox{if \ } h_i = 1 \, , \\
f_0(z_i) & \mbox{if \ } h_i = 0 \\
\end{array}
\right. \\
\mbox{P}(h_i = 1) &=&  \frac{1}{1+\exp\{ -x_i^T \beta \}} \, .
\end{eqnarray*}
Marginalizing out each $h_i$ clearly recovers the original model (\ref{eqn:fdrr1}).  The complete-data log-likelihood for $\beta$ is
$$
l(\beta) = \sum_{i=1}^n \left\{ h_i x_i^T \beta - \log \left(1 + e^{x_i^T \beta} \right) \right\} \, .
$$
This is a smooth, concave function of $\beta$ whose gradient and Hessian matrix are available in closed form.  It is therefore easily maximized using standard methods, such as the Newton--Raphson algorithm.  Moreover, $l(\beta)$ is linear in $h_i$, and the conditional expected value for $h_i$, given $\beta$, is just the conditional probability that $h_i = 1$:
\begin{equation}
w_i = E(h_i \mid \beta, z_i) = \frac{c(x_i) \cdot f_1(z_i)}{ c(x_i) \cdot f_1(z_i) + \{1-c(x_i)\} \cdot f_0(z_i) } \label{bayesoracle3} \, .
\end{equation}
These facts lead to a simple EM algorithm for fitting the model (Algorithm \ref{alg:emPR}; see box).

Thus the overall approach for estimating Model (\ref{eqn:deconvolution1}) has three steps.
\begin{enumerate}[(1)]
\item Fix $\mu$ and $\sigma^2$ under the null hypothesis, or estimate an empirical null (see section \ref{sec:empirical_null}), thereby defining $f_0(z)$.
\item Use predictive recursion to estimate $\pi(\theta)$, and therefore $f_1(z)$, under the two-groups model without covariate effects (see Appendix \ref{ap:PRdetails}).
\item Use $f_0(z)$ and $f_1(z)$ in Algorithm \ref{alg:emPR} to estimate $w_i$ and the regression coefficients.
\end{enumerate}

In principle, the estimate for $\pi(\theta)$ could be improved by using the information in the covariates.  Despite this, our experiments show that the empirical-Bayes approach is essentially just as effective as using a full Bayes approach to estimate $\pi(\theta)$ and the regression function jointly.  This can be explained by the fact that Gaussian deconvolution is such a poorly conditioned problem: for any finite set of observations from $f_1(z)$ in (\ref{eqn:deconvolution1}), there is a large ``near-equivalence'' class of approximately compatible priors $\pi(\theta)$.  Because the Bayes oracle in (\ref{bayesoracle3}) depends on $\pi(\theta)$ only through $f_1(z)$, any prior in this near-equivalence class will yield nearly the same posterior probabilities.  In this sense, predictive recursion seems to provide a ``good enough'' estimate for $\pi(\theta)$, despite ignoring the covariates.

\subsection{Empirical Bayes with marginalization}

\label{sec:FDRregEfron}

In Section \ref{sec:FDRregPR}, we ignored the covariates to estimate $f_1(z)$.  But the most direct generalization of the local-FDR approach of \citet{efrontibshirani2001} would be to ignore the covariates and estimate the overall mixture density $f(z)$ instead.  We now explain why this is a poor solution to the FDR regression problem.  Let us begin with the key insight in \citeauthor{efrontibshirani2001}'s approach to estimating local FDR, which is that the marginal $f(z)$ is common to all test statistics, and that it can be estimated well using the empirical distribution of the $z_i$, without explicit deconvolution of the mixture. This motivates a simple empirical-Bayes strategy: (i) compute a nonparametric estimate $\hat{f}(z)$ of the common marginal density, along with a likelihood- or moment-based estimate $\hat c$ of the mixing fraction; and (ii) plug $\hat f$ and $\hat c$ into the marginalized form of the posterior probability (\ref{bayesoracle2}) to get local FDR for each test statistic.  One caveat is that $\hat c$ must be chosen to ensure that (\ref{bayesoracle2}) falls on the unit interval for all $i$.  But this is an easy constraint to impose during estimation.  See Chapter 5 of \cite{efron:2012} for further details.

We have already remarked that the posterior probabilities in the FDR regression model (\ref{eqn:fdrr1}) do not share a common mixture density $f(z)$, and so cannot be expressed in marginalized form.  Nonetheless, it is natural to wonder what happens if we simply ignore this fact, estimate a global $\hat f(z)$ from the empirical distribution of the test statistics, and use
\begin{equation}
\label{eq:efronFDRR}
1 - w_i^{(t)} =  \frac{\{1-c(x_i)\}   f_0(z_i) }{ \hat f(z_i) }
\end{equation}
in lieu of expression (\ref{bayesoracle3}) used in Algorithm \ref{alg:emPR}.  This has the seemingly desirable feature that it avoids the difficulties of explicit deconvolution.  But (\ref{eq:efronFDRR}) is not guaranteed to lie on the unit interval, and constraining it to do so is much more complicated than in the no-covariates case (\ref{bayesoracle2}).  Moreover, simply truncating (\ref{eq:efronFDRR}) to the unit interval during the course of the estimation procedure leads to very poor answers.

In our simulation studies, we do consider the following ad-hoc modification of (\ref{eq:efronFDRR}), in an attempt to mimic the original local-FDR approach as closely as possible:
\begin{equation}
\label{eq:efronFDRRmodified}
 1- w_i^{(t)} = \left\{
\begin{array}{l l}
1 & \mbox{ if }  (1-\hat c) f_0(z_i)  \geq \hat{f}(z_i)  \, ,\\
T_{u} \left[\frac{\{1-c(x_i)\}   f_0(z_i) }{ \hat f(z_i) } \right]& \mbox{ otherwise. } 
 \end{array}
 \right.
\end{equation}
Here $T_{u}(a)$ is the projection of $a$ to the unit interval, while $\hat{f}(z)$ and $\hat c$ are plug-in estimates of the marginal density and the mixing fraction using the no-covariates method described in Chapter 5 of \cite{efron:2012}.  In our simulation studies, this modification (despite no longer being a valid EM algorithm) does give stable answers with qualitatively correct covariate effects.  But because it zeroes out the posterior probabilities for all $z_i$ within a neighborhood of the origin, it yields heavily biased estimates for $\beta$, and in our studies, it is less powerful than the method of Section \ref{sec:FDRregPR}.

 \subsection{Full Bayes}
 
 \label{sec:FDRregBayes}

From a Bayesian perspective, the hierarchical model
\begin{eqnarray}
(z_i \mid \theta_i) &\sim& \N(\mu + \theta_i, \sigma^2) \nonumber \\
(\theta_i \mid h_i) &\sim& h_i \cdot \pi(\theta_i) + \{1-h_i \}  \cdot \delta_0 \nonumber \\
\mbox{P}(h_i = 1) = c(x_i) &=& \frac{1}{1+\exp(-x_i^T \beta)} \label{eqn:fullbayes_FDRR} \, ,
\end{eqnarray} 
together with priors for $\beta$ and the unknown distribution $\pi(\theta)$, defines a joint posterior distribution over all model parameters.  We use a Markov-chain Monte Carlo algorithm to sample from this posterior, drawing iteratively from three complete conditional distributions: for the mixing density $\pi(\theta)$; for the latent binary variables $h_i$ that indicate whether $z_i$ is signal or null; and for the regression coefficients $\beta$.

An important question is how to parameterize $\pi(\theta)$.  In the no-covariates multiple-testing problem, there have been many proposals, including simple parametric families \citep{scottberger06,polson:scott:2009a} and nonparametric priors based on mixtures of Dirichlet processes \citep{domuller2005}.  In principle, any of these methods could be used.  In our analyses, we model $\pi(\theta)$ as a $K$-component mixture of Gaussians with unknown means, variances, and weights.  We choose $K$ via a preliminary run of the EM algorithm for deconvolution mixture models, picking the $K$ that minimizes the Akaike information criterion (AIC).  (In simulations, we found that AIC was slightly better than BIC at recovering $K$ for the deconvolution problem, as distinct from the ordinary density-estimation problem.)

The model's chief computational difficulty is the analytically inconvenient form of the conditional posterior distribution for $\beta$.  The two major issues here are that the response $h_i$ depends non-linearly on the parameters, and that there is no natural conjugate prior to facilitate posterior computation.  These issues are present in all Bayesian analyses of the logit model, and have typically been handled using the Metropolis--Hastings algorithm.  A third issue, particular to our setting, is that the binary event $h_{i}$ is a latent variable, and not actually observed.

We proceed by exploiting the P\'olya-Gamma data-augmentation scheme for binomial models recently proposed by \citet{polson:scott:windle:2012a}.  Let $x_i$ be the vector of covariates for test statistic $i$, including an intercept term.  The complete conditional for $\beta$ depends only $h = \{h_i: i = 1, \ldots, n\}$, and may be written as
$$
p(\beta \mid h) \propto p(\beta) \prod_{i=1}^n \frac{(e^{x_i^T \beta})^{h_i}}{1+e^{x_i^T \beta} } \, ,
$$
where $p(\beta)$ is the prior.  By introducing latent variables $\omega_i \sim \mbox{PG}(1,0)$, each having a standard P\'olya-Gamma distribution, we may re-express each term in the above product as the marginal of a more convenient joint density:
$$
\frac{(e^{x_i^T \beta})^{h_i}}{1+e^{x_i^T \beta} } \propto e^{\kappa_i x_i^T \beta} \int_0^{\infty} e^{-\omega_i (x_i^T\beta)^2 / 2} \  p(\omega_i) \ d \omega_i \, ,
$$
where $\kappa_i = h_i - 1/2$ and $p(\omega)$ is the density of a PG(1,0) variate.  Assuming a normal prior $\beta \sim \N(c, D)$, it can be shown that $\beta$ has a conditionally Gaussian distribution, given the diagonal matrix $\Omega = \mbox{diag}(\omega_1, \ldots, \omega_n)$.  Moreover, the conditional for each $\omega_i$, given $\beta$, is also in the P\'olya-Gamma family, and may be efficiently simulated.

Together with standard results on mixture models, the P\'olya-Gamma scheme leads to a simple, efficient Gibbs sampler for the fully Bayesian FDR regression model.  Further details of the Bayesian method can be found in Appendix \ref{app:FDRgibbs}, including the priors we use, the conditionals needed for sampling, and the default settings implemented in \verb|FDRreg|.

On both simulated and real data sets, we have observed that the empirical-Bayes and fully Bayes approaches give very similar answers for the local false-discovery rates, and thus reach similar conclusions about which cases are significant.  The advantage of the Bayesian approach is that it provides a natural way to quantify uncertainty about the regression function $s(x)$ and $\pi(\theta)$ jointly.  This is counterbalanced by the additional computational complexity of the fully Bayesian method.

\subsection{Using an empirical null}

\label{sec:empirical_null}

The FDR regression model (\ref{eqn:deconvolution1}) assumes that $\mu$ and $\sigma^2$ are both known, or can be derived from the distributional theory of the test statistic in question.  As \citet{efron:2004} observes, however, many data sets are poorly described by this ``theoretical null,'' and an ``empirical null'' must be estimated instead.  This is a common situation in high-dimensional screening problems, where correlation among the test statistics, along with many other factors, can invalidate the theoretical null.  \citet{efron:2004} proposes two methods for estimating $\mu$ and $\sigma^2$ from the data: (1) maximum likelihood, and (2) central matching, whereby a quadratic function is fit to the log density of some central fraction (e.g.~a third) of the data.

The estimation of an empirical null hypothesis, whether by maximum likelihood or central matching, can be incorporated into the empirical-Bayes method of Section \ref{sec:FDRregPR} as a simple pre-processing step. This approach is used later in our analysis of the neural synchrony data, and is offered as an option in \verb|FDRreg|.  For reasons of identifiability, it is challenging to incorporate an empirical null in a fully Bayesian manner, especially without a strong prior about the null hypothesis.  We refer the reader to \citet{martin:tokdar:2012} for a detailed discussion of this issue, as well as an alternate proposal for estimating an empirical null.  When we use an empirical null in the context of the fully Bayes model described in Section \ref{sec:FDRregBayes}, $\mu$ and $\sigma^2$ are always pre-computed in empirical-Bayes fashion using Efron's method.  This may be slightly less efficient than performing a full Bayesian analysis, but we believe the improved stability is worth the trade-off.

\section{Simulations}

\label{sec:simstudy}

This section presents the results of a simulation study that confirms the advantage of false-discovery-rate regression in problems where signals cluster in covariate space.  We simulated data sets having two covariates $x_i = (x_{i1}, x_{i2})$, with each test statistic $z_i$ drawn according to the covariate-dependent mixture model (\ref{eqn:deconvolution1}).  We considered five choices for $s(x)$:
\vspace{\baselineskip}
\begin{compactenum}[(A)]
\item $s(x) = -3 + 1.5x_1 + 1.5x_2$
\item $s(x) = -3.25 + 3.5x_1^2 - 3.5x_2 ^2$
\item $s(x) = - 1.5(x_1-0.5)^2 - 5|x_2| $
\item $s(x) = -4.25 + 2x_1^2 + 2x_2^2 - 2x_1  x_2$
\item $s(x) = -3$
\end{compactenum}
\vspace{\baselineskip}

\begin{figure}
\begin{tabular}{p{3in} p{3in}}
\includegraphics[width=2.9in]{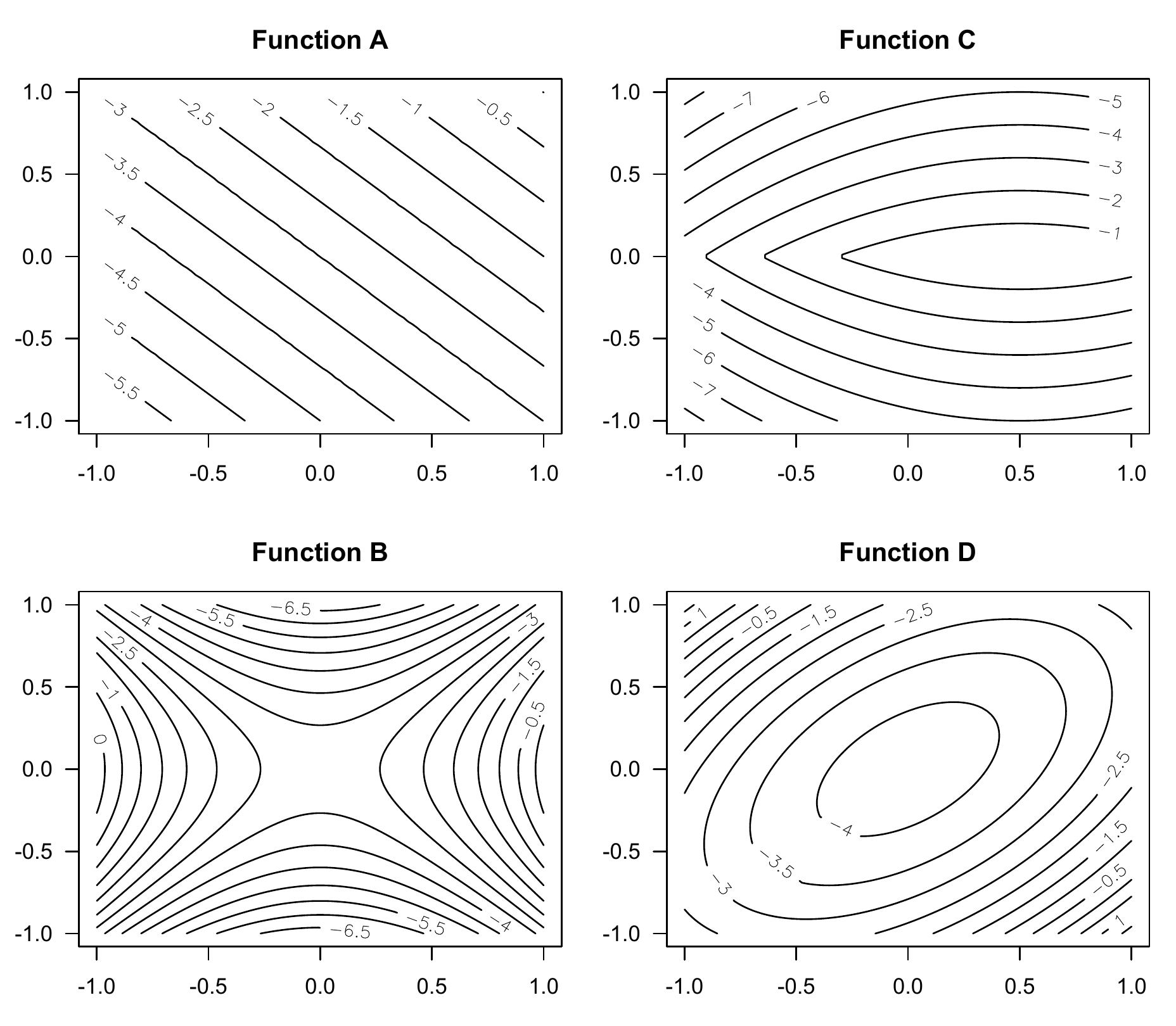} &
\includegraphics[width=2.9in]{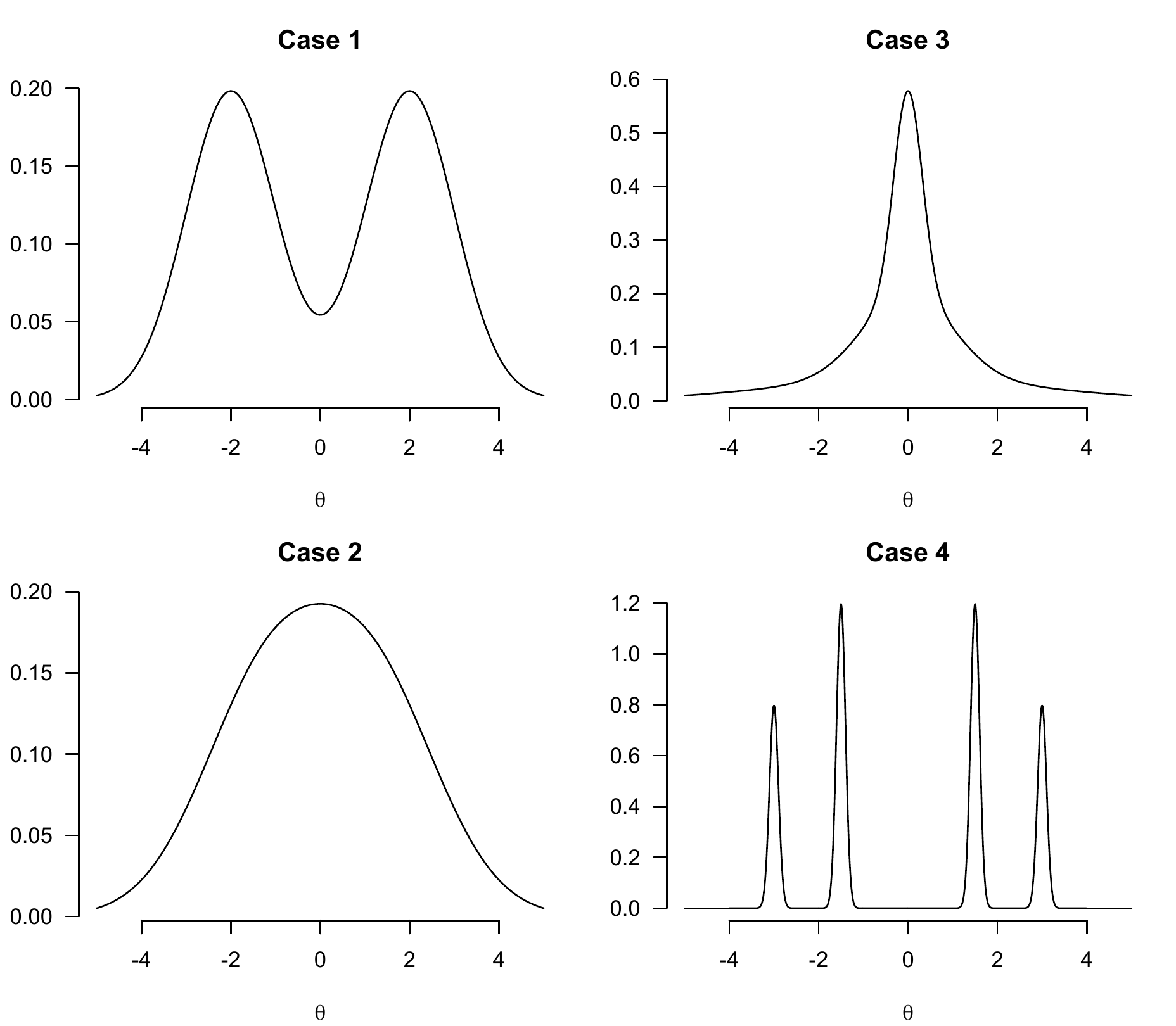} 
\end{tabular}
\caption{\label{fig:simparameters}Settings for the simulation study. Left four panels: choices for the bivariate regression function $s(x_1, x_2)$.  The contours show the prior log odds that a test statistic in that part of covariate space will be a signal.  Function E, not shown, is a flat function: $s(x) = -3$. Right four panels: choices for $\pi(\theta)$.  }
\end{figure}

These choices are shown in the left four panels of Figure \ref{fig:simparameters}.  Function $A$ is linear; Functions B and C are nonlinear but additive in $x_1$ and $x_2$; Function D is neither linear nor additive.  Function E, not shown, is the flat function $s(x) = -3$.  This is included in order to understand the behavior of FDR regression when it is inappropriately applied to a data set with no covariate effects.  The parameters for each function were chosen so that between 6\% and 10\% of the $z_i$ were drawn from the non-null signal population $f_1(z)$.

We also considered four choices for $\pi(\theta)$, all discrete mixtures of Gaussians $\N(\mu, \tau^2)$:

\vspace{\baselineskip}
\begin{compactenum}[(1)]
\item $\pi(\theta) = 0.48 \cdot \N(-2, 1) + 0.04 \cdot \N(0, 16) + 0.48 \cdot \N(2, 1)$
\item $\pi(\theta) = 0.4 \cdot \N(-1.25, 2) + 0.2 \cdot \N(0, 4) + 0.4 \cdot \N(1.25, 2)$
\item $\pi(\theta) = 0.3 \cdot \N(0, 0.1) + 0.4 \cdot \N(0, 1) + 0.3 \cdot \N(0, 9)$
\item $\pi(\theta) = 0.2 \cdot \N(-3, 0.01) + 0.3 \cdot \N(-1.5, 0.01) + 0.3 \cdot \N(1.5, 0.01) + 0.2 \cdot \N(3, 0.01)$
\end{compactenum}
\vspace{\baselineskip}

These choices for $\pi(\theta)$ are shown in the right four panels.  Choices 1 and 4 have most of the non-null signals separated from zero, and are thus easier problems overall. Choices 2 and 3 have most of the signals near zero, and are thus harder problems overall.


For each of the 20 possible combinations for $s(x)$ and $\pi(\theta)$ listed above, we simulated 100 data sets of $n=10000$ test statistics.  Each design point $x_i$ was drawn uniformly from the unit cube, $[0,1]^2$.  To each simulated data set, we applied the following methods.

\vspace{\baselineskip}
\begin{compactdesc}
\item[BH:] the Benjamini--Hochberg procedure \citep{benjamini1995}.
\item[2G:] the two-groups model described in Chapter 5 of \citet{efron:2012}.
\item[EB:] Empirical Bayes FDR regression using predictive recursion (Section \ref{sec:FDRregPR}).
\item[EBm:] Empirical Bayes FDR regression with ad-hoc marginalization (Section \ref{sec:FDRregEfron}).
\item[FB:] Fully Bayes FDR regression using P\'olya-Gamma data-augmentation (Section \ref{sec:FDRregBayes}).
\end{compactdesc}
\vspace{\baselineskip}

For the EB, EBm, and FB methods, we fit a nonlinear additive model by expanding each covariate in a B-spline basis with five equally spaced knots.   Because additivity is assumed, these methods cannot recover Function D exactly, making this a useful robustness check.  The theoretical $\N(0,1)$ null was assumed in all cases.  All methods are implemented in the R package \verb|FDRreg|, and the R script used for the study is available as a supplemental file to the manuscript.

In each case, we selected a set of non-null signals by attempting to control the global false discovery rate at $10\%$ using each method.  We then calculated both the realized false discovery rate and the realized true positive rate (TPR) by comparing the selected test statistics to the truth.  Recall that the true positive rate is defined to be the number of true signals discovered, as a fraction of the total number of true signals.

\begin{table}
\centering
\begin{footnotesize}
\begin{tabular}{r r p{1pc} rrrrr p{1pc} rrrrr}
  \toprule
  & && \multicolumn{5}{c}{False discovery rate (\%)} & &  \multicolumn{5}{c}{True positive rate  (\%)} \\
$\pi(\theta)$ & $s(x)$ && BH & 2G & EBm & EB & FB &  & BH & 2G & EBm & EB & FB \\ 
  \midrule
 1 & A && 8.9 & 9.0 & 9.2 & 9.7 & \signif 10.7 &  & 22.5 & 22.4 & 24.0 & 30.1 & 31.0 \\ 
    & B && 9.5 & 9.5 & 9.4 & 9.7 & \signif 11.1 &  & 21.8 & 21.7 & 23.7 & 32.5 & 34.2 \\ 
    & C && 9.5 & 9.5 & 9.4 & 9.5 & 10.2 &  & 22.8 & 22.4 & 24.2 & 33.3 & 34.3 \\ 
    & D && 9.3 & 9.3 & 9.5 & 9.7 & \signif 11.2 &  & 22.3 & 22.0 & 23.7 & 29.2 & 30.4 \\ 
    & E && 9.4 & 9.0 & 9.7 & \signif 11.0 & 10.2 &  & 18.0 & 17.4 & 18.1 & 18.7 & 18.0 \\ 
   \midrule
   2 & A && 9.2 & 9.1 & 9.3 & 9.7 & 10.2 &  & 13.5 & 13.4 & 14.0 & 18.0 & 18.6 \\ 
    & B && 8.6 & 8.8 & 8.7 & 9.2 & 10.4 &  & 13.0 & 13.1 & 13.7 & 19.0 & 20.2 \\ 
    & C && 9.2 & 9.3 & 9.4 & 9.3 & 10.3 &  & 13.7 & 13.6 & 14.3 & 19.9 & 20.9 \\ 
    & D && 9.3 & 9.3 & 9.5 & 9.7 & 10.6 &  & 13.8 & 13.7 & 14.4 & 17.7 & 18.3 \\ 
    & E && 9.6 & 8.8 & 9.6 & \signif 10.8 & 9.3 &  & 11.3 & 10.9 & 11.3 & 11.7 & 11.1 \\ 
   \midrule
   3 & A && 8.9 & \signif 10.6 & 9.4 & 9.0 & 8.6 &  & 9.4 & 9.7 & 9.7 & 11.1 & 11.0 \\ 
    & B && 9.0 & 10.4 & 9.5 & 8.7 & 8.9 &  & 9.3 & 9.6 & 9.5 & 11.6 & 11.8 \\ 
    & C && 8.6 & 10.0 & 9.0 & 8.4 & 7.9 &  & 9.6 & 9.9 & 9.8 & 11.8 & 11.7 \\ 
    & D && 9.2 & \signif 10.7 & 9.8 & 9.3 & 9.0 &  & 9.8 & 10.0 & 10.0 & 11.3 & 11.2 \\ 
    & E && 10.0 & \signif 10.9 &  10.5 & \signif 11.3 & 9.3 &  & 8.7 & 8.7 & 8.6 & 8.7 & 8.4 \\ 
   \midrule
   4 & A && 9.0 & 9.1 & 9.1 & 10.3 & \signif 10.9 &  & 21.8 & 22.0 & 23.7 & 30.8 & 31.6 \\ 
    & B && 9.4 & 9.4 & 9.4 & 10.3 & \signif 11.1 &  & 21.8 & 21.7 & 24.1 & 33.8 & 34.7 \\ 
    & C && 9.2 & 9.5 & 9.6 & 10.1 & \signif 10.4 &  & 22.5 & 22.7 & 24.6 & 34.7 & 35.2 \\ 
    & D && 9.0 & 9.3 & 9.4 & 9.9 & \signif 11.1 &  & 22.3 & 22.5 & 24.1 & 30.3 & 31.1 \\ 
    & E && 9.9 & 9.3 &  10.1 & \signif 10.9 & 10.0 &  & 17.0 & 16.2 & 17.0 & 17.5 & 16.7 \\ 
   \bottomrule
\end{tabular}
\caption{ \label{tab:simresults} Results of the simulation study.   The rows show different configurations for $\pi(\theta)$ and $s(x)$; see Figure \ref{fig:simparameters}.  The columns show the realized false discovery rate and true positive rate for the five different procedures listed in Section \ref{sec:simstudy}.  The rates are shown as percentages, with results averaged over 100 simulated data sets.  In all cases, the nominal FDR was controlled at the 10\% level.  FDR entries marked with a star are significantly larger than the nominal level of 10\%, as judged by a one-sided $t$-test ($p < 0.05)$.}
\end{footnotesize}
\end{table}

\begin{figure}[t]
\begin{center}
\includegraphics[width=6.0in]{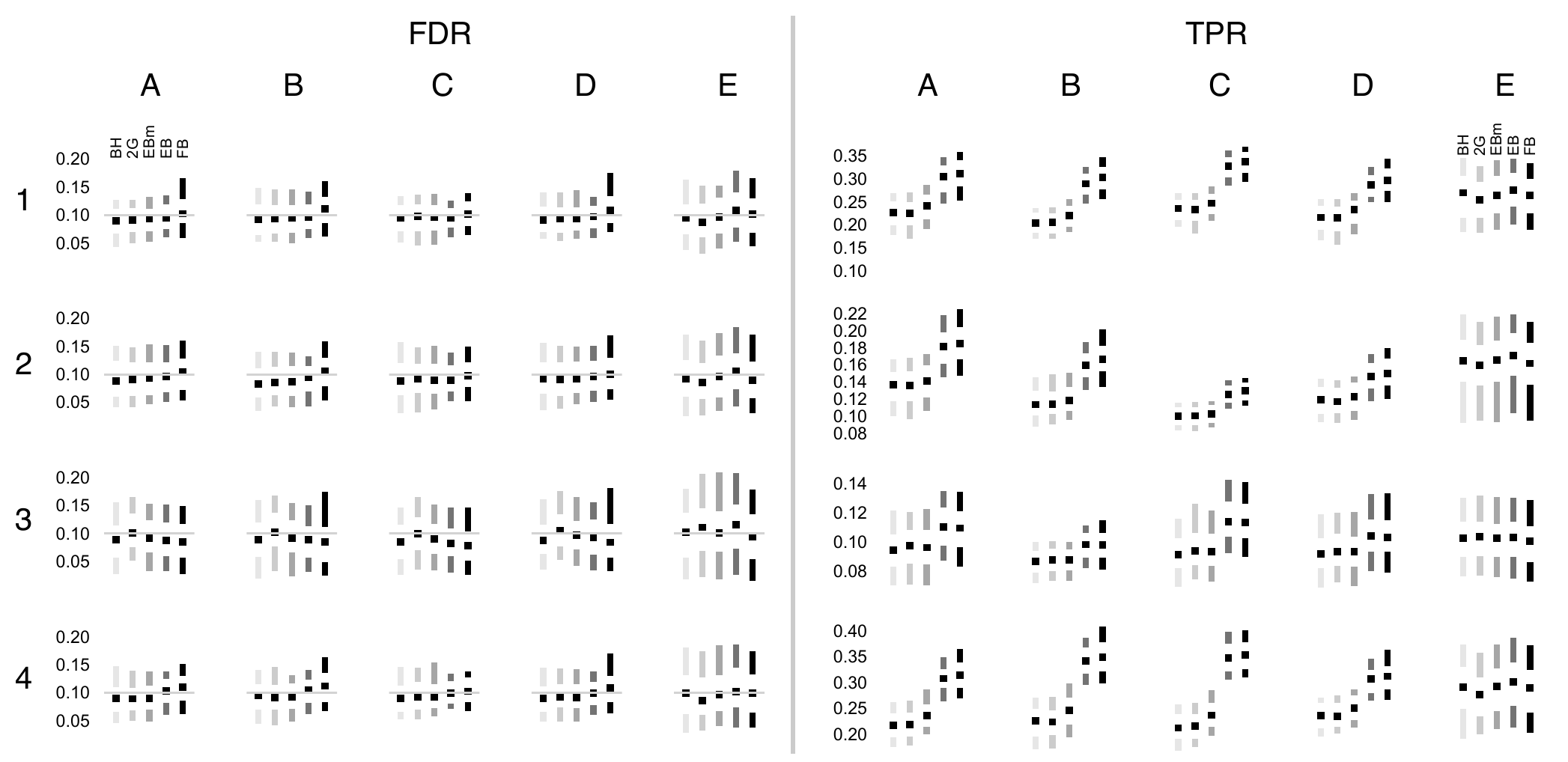}
\caption{\label{fig:sim_results}
Boxplots of false discovery rate (FDR, left 20 panels) and true positive rate (TPR, right 20 panels) for the simulation study.  The rows are the different true priors $\pi(\theta)$, and the columns are the different true regression function $s(x_1, x_2)$.  Within each panel, the boxplots show the median, interquartile range (white bar), and range (grey bar) for each method across the 100 simulated data sets.  Within each panel, the methods are arranged from left to right as: Benjamini--Hochberg (BH), the two-groups model without covariates (2G), empirical Bayes with marginalization (EBm, Section \ref{sec:FDRregEfron}), empirical Bayes (EB, Section \ref{sec:FDRregPR}), and fully Bayes (FB, Section \ref{sec:FDRregBayes}).}
\end{center}
\end{figure}

The results, shown in Table \ref{tab:simresults} and Figure \ref{fig:sim_results}, support several conclusions.  First, the FDR regression method (EB, EBm, and FB) effectively control the false discovery rate at the nominal level (in this case 10\%).  When covariate effects are present, the empirical-Bayes method violates the nominal level about as often as the other methods.   It does so with greater frequency only when covariate effects are absent (function E).  Even then, the average FDR across the different simulated data sets is only slightly higher than the nominal level (e.g. 11\% versus 10\%).  To put this average FDR in context, the realized FDR of the Benjamini--Hochberg method for a single simulated data set typically ranges between 5\% and 15\%.  Therefore, any bias in the regression-based method is small, compared to the variance of realized FDR across all procedures.  (See the left 20 panels of Figure \ref{fig:sim_results}.)

When covariate effects are present, FDR regression has better power than existing methods for a fixed level of desired FDR control.  The amount of improvement depends on the situation.  For priors 1 and 4, the improvement was substantial: usually between 40--50\% in relative terms, or 8--12\% in absolute terms).  For priors 2 and 3, the power gains of FDR regression were more modest, but still noticeable. These broad trends were consistent across the different functions.

The empirical-Bayes and fully Bayes methods perform very similarly overall.  The only noticeable difference is that, when covariate effects are absent (Function E), the empirical Bayes method violates the nominal FDR level slightly more often than the fully Bayes method.  We do not understand why this is so, but as the boxplots in Figure \ref{fig:sim_results} show, the effect is quite small in absolute terms.  They also show that the performance of the empirical-Bayes method in these cases is quite similar in this respect to the two-groups model without covariates (labeled 2G in the plots).

\section{Detecting neural synchrony}

\subsection{Background}

The ability to record dozens or hundreds of neural spike trains
simultaneously has posed many new challenges for data analysis in neuroscience \citep{brown04,buzsaki:2004,aertsen:2010,stevenson11a}.  Among these, the problem of
identifying neural interactions has, since the
advent of multi-unit recording, been recognized as centrally
important \citep{Perkel67a}.  Neural interactions may occur on
sub-behavioral timescales, where two neurons may fire
repeatedly within a few milliseconds of each other. It has been
proposed that such fine-timescale synchrony is crucial for binding
visual objects \citep[see][for opposing views]{gray:1999,shadlen:movshon:1999}, enhancing the strength of communication between
groups of neurons \citep{niebur:etal:2002}, and coordinating the
activity of multiple brain regions \citep{fries:2009,saalmann:kastner:2011}.  It has also been argued that the disruption of synchrony may play a role in
cognitive dysfunctions and brain disorders \citep{uhlhaas:etal:2009}.

Furthermore, there is growing recognition that synchrony and
other forms of correlated spiking have an impact on population
coding \citep{Averbeck06} and decoding \citep{Graf11}. The
proposed roles of neural synchrony in numerous computational
processes and models of coding and decoding, combined with the
knowledge that the amount of synchrony can depend on stimulus
identity and strength \citep{Kohn05} as well as the neuronal
separation \citep{smith:kohn:2008}, make it particularly important
that we have effective tools for measuring synchrony and determining
how it varies under different experimental paradigms.

Rigorous statistical detection of synchrony in the activity of two
neurons requires formulation of a statistical model that specifies
the stochastic behavior of the two neural spike trains under the assumption that they are conditionally independent, given some suitable statistics or covariates \citep{harrison:etal:2012}.  When $n$ neural spike trains are
recorded there $N= { n \choose 2}$ null hypotheses to be tested, which
raises the problem of multiplicity.  In the face of this difficulty, a popular way to proceed has been to control the false-discovery rate using the Benjamini-Hochberg procedure \citet{benjamini1995}, combining all $N$ test statistic into a single analysis.  Yet this omnibus approach ignores potentially useful information about the spatial and functional relationships among individual neuron pairs.
We therefore use false discovery rate regression to incorporate these covariates into an investigation of synchrony in the primary visual
cortex (V1).  Specifically, we analyzed data from V1 neurons recorded from an anesthetized monkey in response to
visual stimuli consisting of drifting sinusoidal gratings, i.e., light whose luminance was governed by a sine wave that moved along an axis having a particular orientation. Details of the experiment and recording technique may be found in \citet{kelly:etal:2007}. Drifting gratings are known to drive many V1 neurons to fire at a rate that depends on orientation. Thus, many V1 neurons will have a maximal firing rate for a grating in a particular orientation (which is the same as an orientation rotated by 180 degrees) and a minimal firing rate when the grating orientation is rotated by 90 degrees. For a given neuron, a plot of average firing rate against angle of orientation produces what is known as the neuron's ``tuning curve." Spike trains from 128  neurons were recorded in response to gratings in 98 equally-spaced orientations, across 125 complete replications of the experiment (125 trials). Here analyze data from the first 3 seconds of each 30-second trial. The 128 neurons generated 8,128 pairs, and thus 8,128 tests of synchrony.  
We applied the model in Equation (\ref{eqn:fdrr1}) to examine the way the probability of synchrony $c(x)$ for a pair of neurons depends on two covariates: the 
distance between the neurons and the correlation of their tuning curves (i.e., the Pearson correlation between the two vectors of length 98 that contain average firing rate as a function of orientation).  
The idea is that when neurons are close together, or have similar tuning curves, they may be more likely to share inputs and thus more likely to produce synchronous spikes, compared to the number predicted under conditional independence.
Our analysis, reported below, substantiates the
observation of \citet{smith:kohn:2008} that the probability of 
fine time-scale synchrony for pairs of V1 neurons tends to decrease
with the distance between the two neurons and increase with the magnitude of tuning-curve correlation.

\subsection{Data pre-processing}

Our analysis takes advantage of a recently-developed technique for measuring
synchrony across binned spike trains (where the time bins
are small, such as 5 milliseconds).  For a pair of neurons
labeled $1$ and $2$, we calculate
\begin{equation}
\label{eq:zetahat}
\hat \zeta = \frac{\mbox{number of bins in which both neurons spike}}
{\sum_t \hat P(\mbox{neuron $1$ spikes at $t$} \mid D^{(1)}_t) \cdot
\hat P(\mbox{neuron $2$ spikes at $t$}  \mid D^{(2)}_t)} \, ,
\end{equation}
where $D^{(1)}_t$ and $D^{(2)}_t$ refer to relevant conditioning information for the firing activity of 
neurons $1$ and $2$ at time $t$, and the
sum is over all time bins across all experimental trials \citep{kelly:kass:2012}. 
The denominator of (\ref{eq:zetahat}) is an estimate of the
number of joint spikes that would be expected, by chance, if $D_t^{(j)}$ 
characterized the spiking activity of neuron $j$ (with $j=1,2$) and,
apart from these background effects, the neurons were independent. 
When $\hat \zeta \approx 1$, or $\log \hat \zeta
\approx 0$, the conclusion would be that the number of observed
synchronous spikes is consistent with the prediction of synchronous
spiking under independence, given the background effects $D^{(j)}$.
Note that this conditioning information $D_t^{(j)}$ 
is intended to capture effects---including tuning curve information---on each neuron separately, whereas the covariates 
that enter the FDRR model (\ref{eqn:fdrr1}) operate pairwise.  Thus, for example, two independent neurons having similar tuning curves would both be driven to fire more rapidly by a grating stimulus in a particular orientation, and would therefore be likely to produce more synchronous spikes by chance than a pair of independent neurons with dissimilar tuning curves. As another example of conditioning information, under anesthesia there are pronounced periods during which most recorded neurons increase their firing rate \citep{brown:etal:2010}.  These waves of increased network activity have much lower frequency than many other physiological wave-like neural behaviors, and are called ``slow waves." One would expect slow-wave activity to account for considerable synchrony, even if, conditionally on the slow-wave activity, a pair of neurons were independent. 
The statistic $\hat \zeta$ in
formula (\ref{eq:zetahat}) is a maximum-likelihood
estimator in the continuous-time framework discussed by \citet{kass:kelly:loh:2011}, 
specifically their equation (22). The purpose of  that framework, and of (\ref{eq:zetahat}), is to describe the way
synchrony might depend on background information. For example,
\citet{kass:kelly:loh:2011} contrasted results from two pairs of
V1 neurons. Both pairs exhibited highly statistically significantly enhanced
synchrony above that predicted by stimulus effects (tuning curves) alone. 
However, the two pairs were
very different with respect to the relationship of synchrony to
slow-wave network activity. In one pair, when background information
characterizing the presence of slow-wave network activity was used in (\ref{eq:zetahat}), the
enhanced synchrony vanished, with $\log \hat \zeta_{1} = .06
\pm .15$.  In the other pair, it persisted with $\log \hat
\zeta_{2} = .82 \pm .23$, indicating the number of
synchronous spikes was more than double the number predicted
by slow-wave network activity together with trial-averaged firing
rate.

The set of 8,128 $\log \hat \zeta$ coefficients analyzed here, together with their standard errors, were created with a model that differed in two ways from that used by \citet{kass:kelly:loh:2011}. First, to better capture slow-wave network effects, in place of a linear model based on single count variable (for neurons $i$ and $j$, Kass et al. used the total number of spikes within the past 50 ms time among all neurons other than neurons $i$ and $j$) a general nonparametric function of the count was fitted, using splines. Second, a nonparametric function capturing spike history effects was used. This allows non-Poisson variability, which is important in many contexts \citep{kass:eden:brown:2014}, and exploratory analysis indicated that it is consequential for this data set as well.

\section{Analysis and results}

As an illustration of the method, we apply false discovery rate regression to search for evidence of enhanced synchrony in a three-second window of recordings on these 128 V1 neurons.  We emphasize that what we refer to here as ``findings'' or ``discoveries'' are necessarily tentative.  As in many genomics data-analysis pipelines, additional follow-up work is clearly necessary to verify any individual discovery arising from an FDR-controlling analysis of a large-scale screening experiment.  Nonetheless, because of its clear covariate effects, the V1 neural recordings provide a good illustration of the FDR regression method.

We use the subscript $i$ to index a pair of neurons, in order to maintain notational consistency with the rest of the paper.  Let $y_{i} = \log \hat \zeta_{i}$ denote the observed synchrony statistic for the $i$th pair of neurons being tested.  This comes from Formula (\ref{eq:zetahat}), after conditioning on slow-wave activity.   Let $s_{i}$ denote the estimated standard error for $ \log \hat{\zeta}_i$, which is obtained from a parametric bootstrap procedure, following \citet{kass:kelly:loh:2011} and \citet{kelly:kass:2012}.  We define $z_i = y_i/s_i$ as our test statistic, and assume that the $z_i$ arise from Model (\ref{eqn:deconvolution1}).  The pairs where $\theta_i = 0$ correspond to the null hypothesis of conditional independence, given slow-wave network activity.  As previously mentioned, there are two relevant covariates: (1) inter-neuron distance, measured in micrometers; and (2) tuning-curve correlation $(r_i)$.

\begin{figure}
\begin{center}
\includegraphics[width=6.0in]{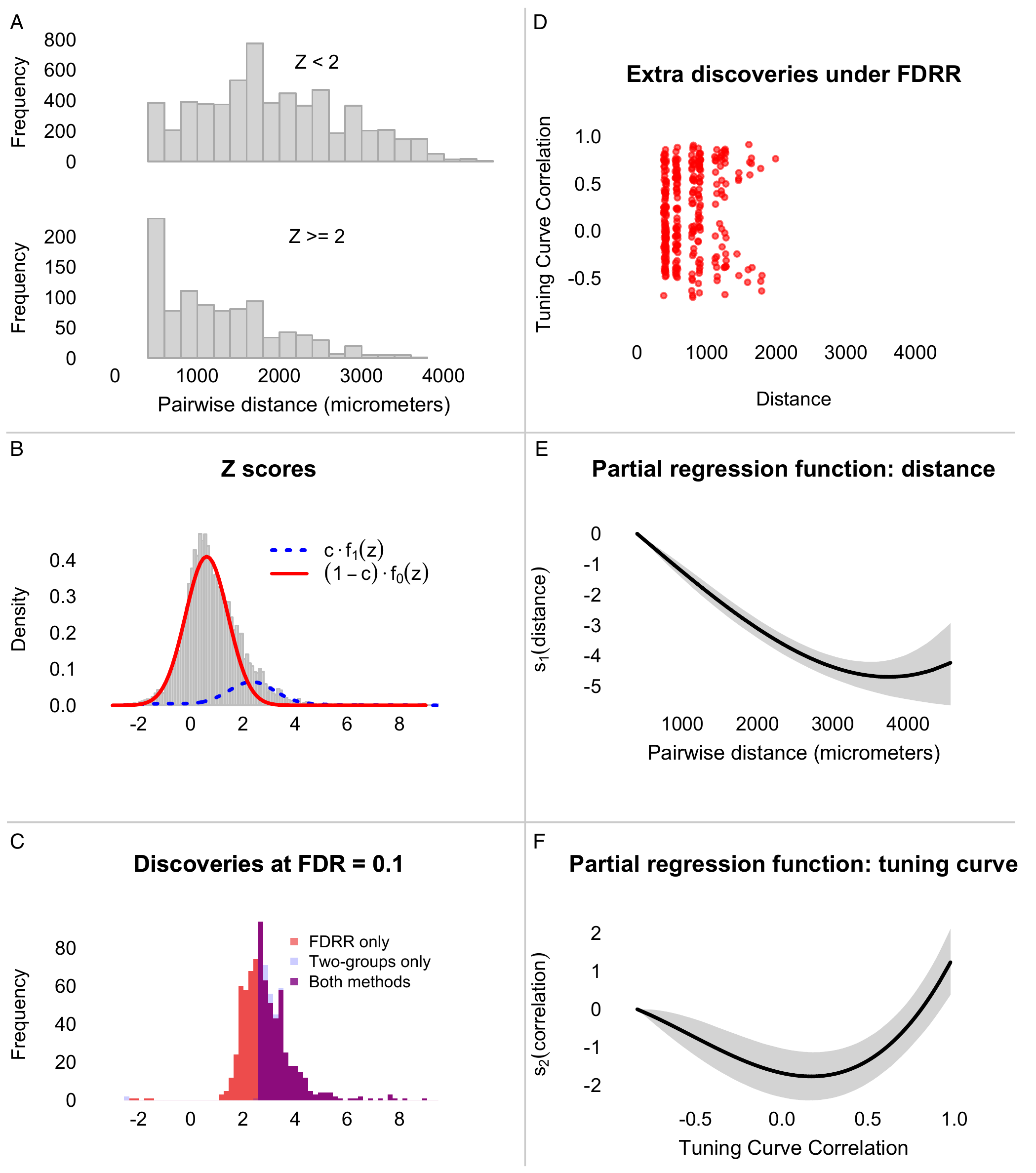}
\caption{\label{fig:synchrony_results}
Panel A shows histograms of inter-neuron distances for pairs with synchrony $z$-score less than 2, versus those with $z$-score larger than 2.  Panel B shows the empirical null density $f_0(z)$, together with the signal density $f_1(z)$ estimated by predictive recursion, superimposed upon the histogram of the raw $z$-scores.  Panel C compares discoveries at the 10\% FDR level using the ordinary two-groups local FDR model, versus those under the FDR regression model.  Panel D shows that the extra discoveries made by FDR regression (red points) tend to concentrate at short inter-neuron distances compared with the rest of the neuron pairs. Panels E and F show the estimated partial regression functions for prior log-odds of being a signal versus distance and tuning curve correlation. The black lines are the estimates, and the grey areas show 95\% posterior credible intervals arising from the full Bayes analysis.}
\end{center}
\end{figure}

Panel A of Figure~\ref{fig:synchrony_results} provides some initial exploratory evidence for a substantial distance effect.  It shows two histograms of $z$-scores: one for neuron pairs where $z_i < 2$ (suggesting no synchrony enhancement), and another for neuron pairs where $z_i \geq 2$ (suggesting possible synchrony enhancement).  It is clear from the figure that nearby neuron pairs are much more likely to have $z_i \geq 2$, versus neuron pairs at a longer distance.  This motivates the use of covariate-dependent prior probabilities in (\ref{eqn:deconvolution1}).

To fit the FDR regression model, we first estimated an empirical null $f_0(z)$, as described in Section \ref{sec:empirical_null}.  This was necessary because the empirical distribution of $z$-scores was poorly described by a standard normal density.  We used the maximum-likelihood method from \citet{efron:2004}, which yielded $\mu = 0.61$ and $\sigma = 0.81$.  This suggested underdispersion and a positive bias versus the theoretical $\N(0,1)$ null.  Fixing $\mu$ and $\sigma$ at these estimated values, we then ran predictive recursion to estimate $f_1(z)$, as described in Section \ref{sec:FDRregPR}.  Panel B of Figure \ref{fig:synchrony_results} shows the estimates for $f_0(z)$ (solid red line) and $f_1(z)$ (dashed blue line), scaled by the empirical-Bayes estimate of the mixing fraction $c$ in the two-groups model (\ref{basicmixturemodel}), and superimposed on the histogram of $z$-scores.  The alternative hypothesis appears to be dominated by cases where $z_i > 0$.

Having computed estimates for $f_0(z)$ and $f_1(z)$, we then used the empirical-Bayes method of Section \ref{sec:FDRregPR} to estimate the FDR regression model by expectation--maximization.   We assumed that the prior log odds of synchrony ($\theta_i \neq 0$) could be described by an additive model involving distance and tuning-curve correlation:
$$
s(x) = \beta_0 + s_1(\mathrm{distance}) + s_2(\mathrm{correlation}) \, .
$$
The partial regression functions were modeled by expanding each covariate in a B-spline basis, with the degrees of freedom chosen to minimize AIC.  To regularize the estimates, we used $N(0,1)$ priors on the spline coefficients.  The partial regression functions are identified only up to additive constants.  To identify them, we estimated an overall intercept $\beta_0$, and fixed $s_1$ and $s_2$ to be zero at their left-most endpoints.  As a robustness check, we also ran the full Bayes method, which does not require a pre-computed estimate for $f_1(z)$.  We focus mainly on results for the empirical-Bayes approach, but the fully Bayes estimates of local FDR were very similar, and we use the full Bayes method to construct confidence bands for the underlying regression function.

We controlled the (Bayesian) false discovery rate at the 10\% level, and compared the resulting discoveries under the FDR regression model to those under the ordinary two-groups (local FDR) model without covariate effects.  The regression model yielded roughly 50\% more discoveries compared to the two-groups model, 763 versus 489.  Panels C and D of Figure~\ref{fig:synchrony_results} show that these extra discoveries tend to be at the borderline of statistical significance ($z_i \approx 2$), but heavily concentrated at short distances, where the prior odds of a significant $z$-score are much higher.

Panels E and F of Figure~\ref{fig:synchrony_results} show the estimated partial regression functions $s_1$ and $s_2$, together with 95\% posterior credible intervals derived from the fully Bayesian posterior distribution of the spline coefficients.  The distance effect suggested by Panel A is confirmed by the confidence bands of the partial regression function for distance in Panel E.  Tuning-curve correlation also appears to play a role in the prior odds of synchrony, with its effect roughly (though not exactly) symmetric about zero.  To provide intuition about the magnitude of the covariate effects, we compare two sets of pairs.
\begin{itemize}
\item A neuron pair at distance 2433 micrometers (the 75th percentile), and with tuning-curve correlation of $0.12$ (the median), was estimated to have a $2.3\%$ prior probability of being a non-null signal.  A neuron pair with the same tuning curve correlation but separated by only 1200 micrometers (the 25th percentile) was estimated to have a $16.5\%$ prior probability of being a non-null signal.
\item A neuron pair at distance 1789 micrometers (the median), and with tuning-curve correlation of 0, was estimated to have a $6.9\%$ prior probability of being a non-null signal.  Another neuron pair at the same distance of 1789 micrometers, but with tuning-curve correlation of $0.5$, was estimated to have a $10\%$ prior probability of being a non-null signal.  A third pair at the same distance and tuning-curve correlation of $0.75$ was estimated to have a $24\%$ prior probability of being a non-null signal.
\end{itemize}

\section{Final remarks}

Our FDR regression model preserves the spirit of the unified Bayes/frequentist approach of the two-grouds model (\ref{basicmixturemodel}) while incorporating test-level covariates that, in our motivating example, describe the physical and functional relationships among neurons.  Our results show that involving these covariates directly in the multiple-testing model has the potential to improve inferences about fine-time-scale neural interactions.  While we consider our findings to be preliminary, the distance and tuning-curve effects suggested by our analysis are easily interpretable, and support the previous analyses of \citet{smith:kohn:2008}.

The neural-recordings data set we have analyzed here is typical of many found in today's pressing scientific problems, in that it exhibits two important statistical features: the need to adjust for simultaneous inference, and the presence of spatial information, or some other nontrivial covariate structure.  Previous attempts to handle this structure have typically involved separate analyses on subsets of the data, such as the ``front-versus-back of brain'' split considered by \citet{efron:2008b}.  When there is an obvious subset structure in the data, such an approach may be appealing.  Yet it requires case-by-case judgments, and opens the door to further multiplicity issues regarding the choice of subsets.  Our results show that false-discovery-rate regression can avoid these difficulties, without compromising on the global error rate, by incorporating covariates directly into the testing problem.  It is therefore suited to the increasingly common situation in which test statistics should not be considered exchangeable.

\paragraph{Acknowledgements.}  The authors thank the editor, associate editor, and two anonymous referees for their detailed and helpful feedback.  Scott was partially supported by a CAREER grant from the U.S.~National Science Foundation (DMS-1255187).

\appendix

\section{Predictive recursion}
\label{ap:PRdetails}

Predictive recursion is used to estimate the mixing distribution $\pi(\theta)$ in the following formulation of the two-groups model without covariates:
\begin{eqnarray}
z_i &\sim& c \cdot f_1(z_i) +  (1-c)  \cdot f_0(z_i) \nonumber \\
f_0(z) &=& \N(z \mid \mu, \sigma^2) \nonumber  \\
f_1(z) &=& \int_{\mathcal{R}} \N(z \mid \mu + \theta, \sigma^2) \ \pi(\theta) \label{eqn:nocovars_deconvolution} \, .
\end{eqnarray}
An equivalent formulation is 
\begin{eqnarray}
z_i &\sim& \N(\mu + \theta_i, \sigma^2) \nonumber \\
\theta_i &\sim& \Psi \, , \quad \Psi =  \tilde{\pi}_1(\theta)  + \pi_0 \delta_0 \nonumber \, ,
\end{eqnarray}
where $\Psi$ is absolutely continuous with respect to the dominating measure $\nu$ defined as the sum of Lebesgue measure on $\mathcal{R}$ and a point mass at $0$.  Here $\tilde{\pi}_1(\theta) = c \cdot \pi(\theta)$ is a sub-density corresponding to signals, and $\pi_0 = 1-c$ is the mass at zero corresponding to nulls.

\begin{algorithm}[t]
\KwData{Test statistics $z_1, \ldots, z_n$}
\KwIn{Null model $\N(\mu, \sigma^2)$; weights $\gamma^{[i]}$; initial guess $\Psi^{[0]} = \tilde{\pi}^{[0]}_1(\theta)  + \pi^{[0]}_0 \delta_0$ having a continuous sub-density $\tilde{\pi}^{[0]}_1(\theta)$ and a Dirac measure at zero of mass $\pi_0^{[0]}$.}
\For{$i=1, \ldots, n$}{
\begin{eqnarray*}
m_0^{[i]} &=& \pi_0^{[i-1]} \cdot \N(z_i \mid \mu, \sigma^2) \\
f_1^{[i]}(\theta) &=& \N(z_i \mid \mu + \theta, \sigma^2) \ \tilde{\pi}_1^{[i-1]} (\theta) \quad (\mbox{discrete grid}) \\
m_1^{[i]} &=& \int_\mathcal{R} f_1^{[i]}(\theta)  \ d \theta \quad \quad \mbox{(trapezoid rule)} \\
\pi_0^{[i]} &=& (1-\gamma^{[i]}) \cdot \pi_0^{[i-1]} + \gamma^{[i]} \cdot \left( \frac{m_0^{[i]} }{m_0^{[i]}  + m_1^{[i]} } \right) \\
\tilde{\pi}^{[i]}_1(\theta) &=&  (1-\gamma^{[i]}) \cdot\tilde{\pi}^{[i-1]}_1(\theta)  + \gamma^{[i]} \cdot \left( \frac{f_1^{[i]}(\theta) }{m_0^{[i]}  + m_1^{[i]} } \right) 
\end{eqnarray*}
}
\KwOut{Estimates $c = 1-\pi_0^{[n]} $ and $\pi(\theta) = \pi_{1}^{[n]}(\theta) / c$. 
}
\caption{\label{alg:PR} Predictive recursion for estimating ${\pi}(\theta)$ and $c$ in Model (\ref{eqn:nocovars_deconvolution}).  The subdensity $\tilde{\pi}_1(\theta) = c \pi(\theta)$ is approximated on a discrete grid, and integrals with respect to $\tilde{\pi}_1(\theta)$ are calculated by the trapezoid rule.}
\end{algorithm}

Predictive recursion \citep{newton:2002} is a stochastic algorithm for estimating $\Psi$, or for any mixing density with respect to an arbitrary dominating measure $\nu$, from observations $z_1, \ldots, z_n$.  Assume that $\mu$ and $\sigma^2$ are fixed.  Begin with a guess $\Psi^{[0]}$ and a sequence of weights $\gamma^{[i]} \in (0,1)$.  For $i = 1, 2, \ldots, n$, recursively compute the update
\begin{eqnarray}
m^{[i-1]}(z_i) &=& \int_\mathcal{R} \N(z_i \mid \mu + u, \sigma^2) \  \Psi^{[i-1]}(du) \label{eqn:pr1} \\
\Psi^{[i]}(du) &=& (1 - \gamma^{[i]}) \Psi^{[i-1]} (du) + \gamma^{[i]} \cdot \left\{ \frac{ \N(z_i \mid \mu + u, \sigma^2) \Psi^{[i-1]}(du)}{m^{[i-1]}(z_i)} \right\}  \, . \label{eqn:pr2}
\end{eqnarray}
The final update,
$$
\Psi^{[n]} = \tilde{\pi}^{[n]}_1(\theta)  + \pi^{[n]}_0 \delta_0 = c^{[n]} \cdot \pi^{[n]}(\theta)  + (1-c^{[n]}) \cdot \delta_0 \, ,
$$
provides estimates for $c$ and the mixing density $\pi(\theta)$.  In practice, the continuous component $\pi(\theta)$ is approximated on a discrete grid of points, and the integral in (\ref{eqn:pr1}) is computed using the trapezoid rule over this grid. 

The key advantages of predictive recursion are its speed and its flexibility.  Moreover, \citet{tokdar:martin:ghosh:2009} derive conditions on the weights $\gamma^{[i]}$ that lead to almost-sure weak convergence of the PR estimate to the true mixing distribution.  They also show that, when the mixture model is mis-specified, the final estimate converges in total variation to the mixing density that minimizes the Kullback-Leibler divergence to the truth.  The conditions on the weights $\gamma^{[i]}$ necessary to ensure these results are satisfied by $\gamma^{[i]} = (i+1)^{-a}$, $a \in (2/3, 1)$.  We use the default value $a=0.67$ recommended by \citet{tokdar:martin:ghosh:2009}.

Algorithm \ref{alg:PR} describes the steps of predictive recursion in detail, including a clear separation of the continuous and discrete components of the mixture distribution $\Psi$.  In our implementation, we pass through the data 10 times, randomizing the sweep order in each pass.  This yields stable estimates that are relatively insensitive to the order in which the data points are processed, and is consistent with the practice of other authors who have studied predictive recursion \citep{newton:2002,tokdar:martin:ghosh:2009,martin:tokdar:2012}.


\section{Details of the fully Bayes method}
\label{app:FDRgibbs}

Our implementation of the fully Bayes FDR regression model in (\ref{eqn:fullbayes_FDRR}) assumes that $\pi(\theta)$ is a $K$-component discrete mixture of Gaussians, parametrized by a set of component weights $\eta_k$, means $\mu_k$ and variances $\tau^2_k$.

Let $h_i$ be the binary indicator of whether $z_i$ is signal or noise, let $\beta$ denote the regression vector, and let
$$
c(x_i) = \frac{1}{1+e^{-x_i^T \beta}} \, .
$$
We assume the conditionally conjugate priors
\begin{eqnarray*}
\beta &\sim& N(b_0, B_0) \\
\mu_k &\sim& N(0, v_\mu) \\
\tau^2_k &\sim& IG(a/2, b/2) \\
(\eta_1, \ldots, \eta_K) &\sim& \mbox{Dirichlet}(\alpha) \, .
\end{eqnarray*}
Under these priors, the full conditionals needed to implement a Gibbs sampler are as follows. To lighten the notation, a dash ($\text{---}$) is used to denote ``all variables not otherwise named.''  Our simulation studies use the prior parameters $b_0 = 0$, $B_0 = 100 I$, $v_{\mu} = 100$, $a=b=1$, and $\alpha = (1, \ldots, 1)$.

To update $h_i$, note that, from standard results on Gaussian mixtures, the conditional predictive density under the alternative is
$$
f_1(z_i \mid \text{---}) = \sum_{k=1}^K \N(z_i \mid \mu_k, \tau^2_k + \sigma^2) \, .
$$
Thus to draw $h_i$, we sample from the Bernoulli distribution
$$
(h_i \mid \text{---}) \sim \left\{ 
\begin{array}{l l}
1 & \mbox{with probability $w_i$} \\
0 & \mbox{otherwise,} 
\end{array}
\right.
$$
where
$$
w_i = \frac{c(x_i) \cdot f_1(z_i \mid \text{---})}{c(x_i) \cdot f_1(z_i \mid \text{---}) + \{1-c(x_i)\} \cdot f_0(z_i)} \, .
$$

Conditional upon $h_i$, the regression coefficients $\beta$ can be updated in two stages using the Polya-Gamma latent-variable scheme.  First, draw auxiliary variables $\omega_i$ from a Polya-Gamma distribution as
$$
(\omega_i \mid \text{---}) \sim \mbox{PG}(1, x_i^T \beta) \, ,
$$
using the method of \citet{polson:scott:windle:2012a}, and implemented in \citet{bayeslogit:2013}.  Let $\Omega = \mbox{diag}(\omega_1, \ldots, \omega_n)$ and $\kappa = (h_1 - 1/2, \ldots, h_n - 1/2)$. Use these to update $\beta$ as
$$
(\beta \mid \text{---}) \sim \N(m_\beta, V_\beta) \, ,
$$
where
\begin{eqnarray*}
V_\beta^{-1} &=& X^T \Omega X + B_0^{-1} \\
m_\beta &=& V_\beta^{-1} (X^T \kappa + B_0^{-1} b_0) \, .
\end{eqnarray*}

Given the $z_i$ corresponding to signals ($h_i = 1$), the mixture-model weights, means, and variances involve straightforward conjugate updates, and are described in many standard textbooks on Bayesian analysis.  Thus we do not include them here; see, for example, Chapter 22 of \citet{gelman:carlin:stern:rubin:2004}.


\end{spacing}

\singlespace

\bibliographystyle{abbrvnat}
\bibliography{synchrony}

\end{document}